\documentstyle[epsfig,12pt]{article}
\begin{document}

\begin{center}
{\Large \bf Changes in the Hurst exponent of heartbeat intervals during
physical activities}
\vspace{0.5cm} \\ 
M. Martinis$^{1}$, A. Kne\v zevi\' c$^{1}$, G. Krsta\v ci\' c$^{2}$ and 
E. Vargovi\' c$^{3}$ \\
{$^{1}$\em Rudjer Bo\v skovi\' c Institute, Zagreb, Croatia} \\
{$^{2}$\em Institute for Cardiovascular Disease and Rehabilitation, Zagreb,
Croatia} \\
{$^{3}$\em CDV info, Zagreb, Croatia} 
\end{center} 
 
\vspace*{0.5cm}
\begin{small}
The fractal scaling properties of the heartbeat time series are studied in a
controlled ergometric regime using the Hurst rescaled range R/S analysis.
The long-time "memory effect" quantified by the value of the Hurst exponent $H>0.5$
is found to increase during progresive physical activity at healthy 
subjects in contrast to those having stable angina pectoris (SAP), 
where it is decreasing.
We argue that this finding may be used as a useful new diagnostic 
parameter for short heartbeat time series.
\end{small}

\vspace*{0.5cm}

The output of many physiological systems, such as the normal human 
heartbeat time series, are extremely inhomogeneous and nonstationary. 
They fluctuate in an irregular and complex manner, even under resting 
conditions. 
The presence of scaling properties suggest that the
regulatory systems are operating far from equilibrium and
application of fractal analysis may provide a new approach to recognize
deseased states by studying changes in the scaling properties.  
It has been observed that fractal scaling is degraded in some deseased 
states \cite{ref1,ref2,ref3}.

The heart rate variability (HRV) under controlled physical activity 
is not well studied \cite{ref4,ref5}.
The time series of heartbeat intervals (RR intervals), 
used in various analyses, are usually holter type data or data from 
steady state measurement. 
The conventional statistical analysis of the ambulatory ECG (Holter) 
records is, however, limited.
The great wealth of data about the 
dynamics of the heart that is contained in such records is usually reduced 
to characterize only the mean heart rate and the presence and
frequency of some abnormal electrocardiographic complexes. The analysis
of long-time correlations is largely ignored. One possible reason for this 
is that in the study of twenty four hour HRV, in a natural setting, the subjects
are constantly changing their level of physical activity. 
These changes are unpredictable and the comparison of such time series are
difficult. 
However, some heart deseases, such as stable angina pectoris (SAP), are clearly visible
only under physical activity. 
That has given us a reason for investigating 
the scaling properties of heartbeat intervals in a various well defined 
ergometric regimes.  

Fluctuations in RR intervals during one of our ergometric measurements 
are shown in Fig.1.
Both time series look very similar and we canot say, without knowing,
which one is from a healthy and which one is from a SAP individual.

\begin{figure}
\begin{center}
\epsfig{file=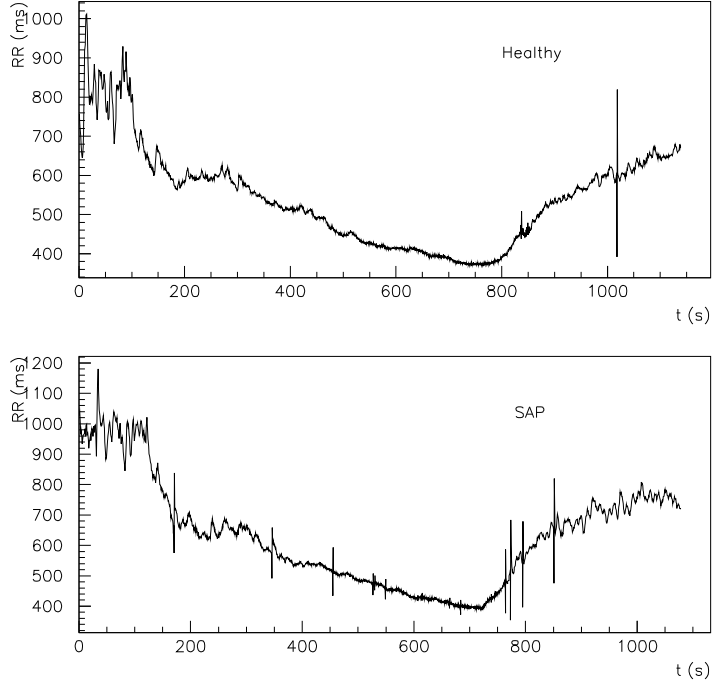,width=10.5cm}
\caption{RR intervals of a healthy (top) and a SAP (bottom) subject 
         in one of our ergometric measurements.} 
\end{center}
\end{figure}

Different self-similar time series can be clasified into random or
non-random series by estimating their Hurst exponent \cite{ref6}.  
In order to determine the long-time dependencies in time series of RR 
intervals during controlled physical activity, we apply the rescaled range
(R/S) method \cite{ref7}, to estimate the Hurst exponent.  
In fact, we are interested in the capability of the R/S method to distinguish 
the patients with SAP from the healthy subjects. 
For a heartbeat time series data of length N \{ $u(n)\, , 
 n=1,\ldots ,N$ \}, where $u(n)\equiv RR(n)=t(R_{n+1})-t(R_{n})$ is the 
$n^{th}$ RR interval defined as a difference in time position for R-wave peaks, 
we calculate the running means ${\overline u(n)}$ for a given $n$ and the 
accumulated deviations from the mean X(l,n), l=1,...,n:
\[ {\overline u(n)} =\frac{1}{n} \sum_{k=1}^{n} u(k), \] 
\[ X(l,n)=\sum_{k=1}^{l} [u(k)-{\overline u(n)}]. \]
The range $R(n)$ is the distance between the minimum and the maximum 
value of $X$, 
 and is rescaled by dividing it by the standard deviation $S(n)$:
\[ R(n)=max_l X(l,n)-min_l X(l,n), \] 
\[ S(n)=\sqrt{\frac{1}{n} \sum_{k=1}^{n} (u(k)-{\overline u(n)})^{2}}. \]
The rescaled range (R/S) is a dimensionless quantity and for large $n$
it is expected to show a power law dependence:
 \[ R(n)/S(n)\sim n^{H}, \]
where H is the Hurst exponent. If the time series is long enough,
the relationship between the fractal dimension D and the Hurst exponent (H) is
\[ D=2-H. \]
The time series of RR intervals can be divided into three distinct
categories: H$<$0.5, H=0.5 and H$>$0.5. The case H=0.5 correspond to random
or uncorrelated RR intervals. If $H>0.5$, the RR intervals are
persistent and characterized by long-time correlations or "memory' effects
on all time scales. The strength of the persistence increases as H
approaches 1.0. The impact of the present on the future can be estimated
through the correlation function (C)
\[ C=2^{(2H-1)} -1. \] 
Most of natural phenomena show persistent behavior with 
H$\sim 0.7$ \cite{ref7,ref8}.
The time series with $H<0.5$ is antipersistent, which means that RR
intervals are negatively correlated.

The time series of RR intervals in our controled ergometric measurement
(Fig.2) had a time duration of about 15 min. ($\simeq$ 2000 beats). 
This type of measurement is used as a routine in everyday clinical
diagnostic procedure, because some heart deseases, such as SAP, 
usually become transparent under physical activities.

The ECG ergometric data were digitized at sampling time of 1 ms 
by the WaveBook 512 (Iotech. Cal. USA), and
transferred to a computer for further analysis. The RR interval series was 
passed through a filter that eliminates noise and artefacts. 
All R-wave peaks were first edited automatically, after which a careful 
manual editing was performed by visual inspection of the each RR interval. 
After this, all questionable portions were excluded manually, 
and only segments with $>90\%$ sinus beats were included in the final analysis.  
The location of the R-wave peaks was determined with a resolution of 1 ms. 

Each measurement consists of stationary state part (pretriger Pt), 
few stages of running (P1-P4) on an inclined belt and a period of 
relaxation (Re). 
Different regimes of physical activity are defined according to the 
standard Bruce protocol (Table 1), with a time duration of 3 min for 
each program. The pretriger part has a variable duration 
and is limited for analysis to the first 30 sec in each measurement. 
The relaxation period is restricted to 6 min.

\begin{figure}
\begin{center}
\epsfig{file=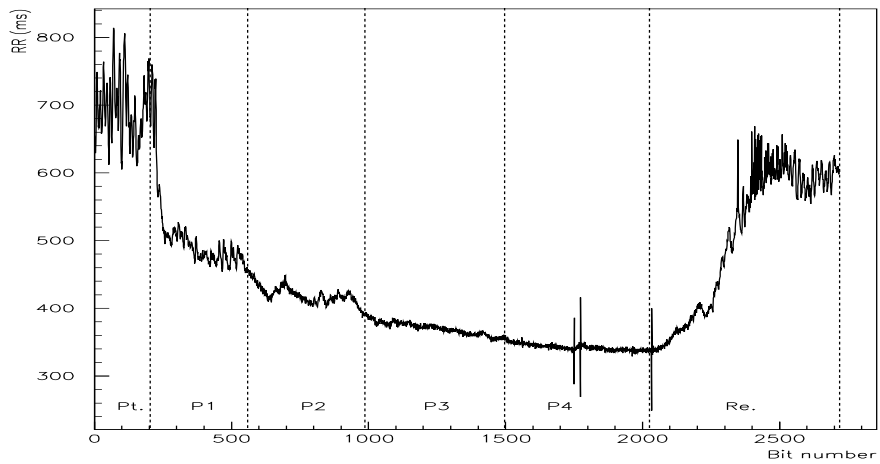,width=13.0cm,height=7cm}
\caption{RR intervals in an ergometric measurement. The global 
         nonstationarity as result of physical activity
         is clearly seen. Pt correspond to stationary state measurement,
         P1-P4 to running stages on a belt with increasing intensity, 
         and Re to a relaxation period after stopping the moving belt.}
\end{center}
\end{figure}

\begin{small}
\begin{table}
\begin{center}
\begin{tabular}{|c|c|c|}
\hline
Program & Belt angle ($^\circ$) & Belt velocity (km/h) \\
\hline
  P1     &  10                  &     2.7 \\
  P2    &  12                  &     4   \\
  P3   &  14                  &     5.5  \\
  P4    &  16                  &     6.9  \\
  P5     &  18                  &     8   \\
%  P6    &  20                  &     8.8 
\hline
\end{tabular}
\caption{Defined regimes in ergometric measurement: Brouce protocol.}
\end{center}
\end{table}
\end{small}

The R/S analysis was performed on four separated regimes:
stationary state Pt, running programs P1, P2 and relaxation Re. 
Patients were divided in two groups: one with the evidence of ishemic
ST-segment depression of more than 1 mV (SAP subjects), and the control group 
of healthy subjects.
Selection of subjects was performed by a cardiologist according to the 
generally accepted medical knowledge.
Before starting with the R/S calculation, we performed a 
polynomial regresion (trendline) fit on the original 
RR interval data, in order to compensate for the global nonstationarity 
in the data caused by the physical activity.   
The R/S calculation was performed on the deviation of the original RR
intervals from the trendline.
The main steps are shown in Fig.3 for P1 regime.
\begin{figure}
\begin{center}
\epsfig{file=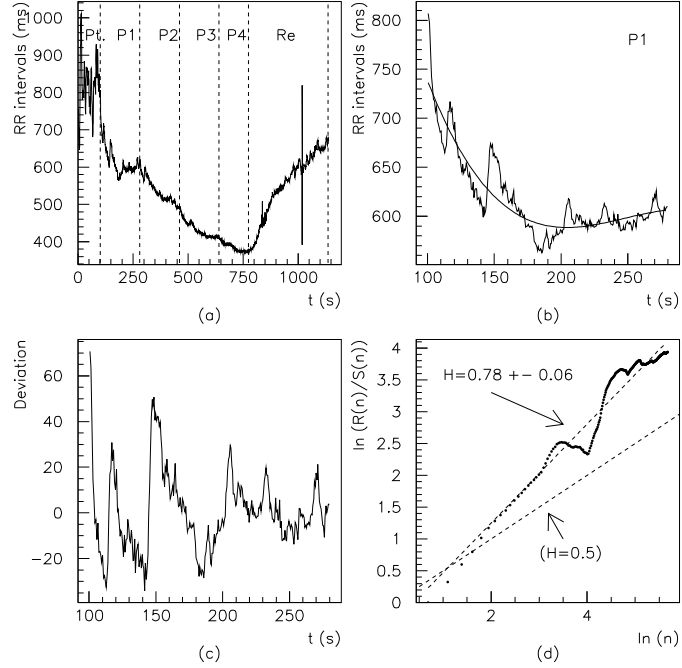,width=10.0cm}
\caption{(a)Typical shape of HRV in the ergometric measurement,
(b) RR intervals in Program 1 with 3$^{rd}$ degree polynomial
regresion trendline,
(c) deviations from the polynomial trendline,
(d) R/S results for deviation signal in comparison with
    random data result (H=0.5).}
\end{center}
\end{figure}
The procedure adopted here is to calculate R/S for a box of 
$n$ elements, starting with the first two elements. In each next step
one more element is added, and R/S is calculated for the wider box.
The process is continued until the box of length N (the whole data set)
is reached.
The Hurst exponent H is evaluted as a slope of the least-square fit line
on $log(R/S)$ vs. $log(n)$ plot. 
In this way, we preserve the ordering of RR intervals during calculation.

The average Hurst exponent H in four analysed regimes, with corresponding
statistical error bars are shown in Fig.4, for healthy and SAP subjects.
It includes 14 independent measurements on 7 healthy + 7 SAP subjects.

\begin{figure}
\begin{center}
\epsfig{file=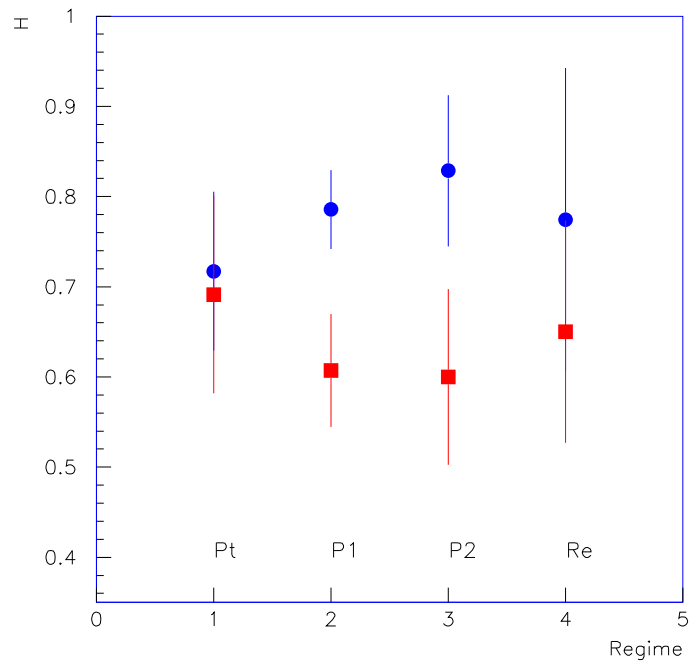,width=13cm}
\caption{Average Hurst exponents with corresponding 1$\sigma$ error 
         bars for different regimes of ergometric measurement; Pretrigger (Pt),
         Program 1 (P1), Program 2 (P2) and Relaxation (Re). 
         Dots are for healthy subjects, and squares are for SAP subjects.}
\end{center}
\end{figure}

The Hurst exponent H exceeds 0.5
in all cases and ranges from 0.6--0.9 depending on the regime type. 
For RR intervals in the Pt regime, H for both healthy and SAP 
subjects is about 0.7 and we cannot distinguish them. 
The situation is different in regimes under physical activity.
The difference between healthy/SAP subjects is clearly seen in the P1--P2
program of controlled running, in our analysis. 
The Hurst exponent for healthy subjects increases with increased 
running intensity, while H for SAP subjects decreases in the
same (P1--P2) regimes. 
Our time series are too short for reliable  connection of the scaling 
exponent with the fractal dimension. However, if the scaling exponent 
reveals the complexity of the underlying pattern, 
we may say that the complexity of the RR pattern in 
running is decreased in the healthy heart, but increased in the sick heart.
The conclusion is opposite from what it would be expected in a stationary state
measurement \cite{ref3}. It could be connected with the heart reaction 
on well defined excitation. 
The healthy heart restricts flexibility according to restrict outside 
condition: well defined excitation -- well defined reaction. The sick 
heart, under well defined excitation, probably produces "mess" in the reaction, 
and a more complex pattern of fluctuation.        

In the relaxation part, which is a recovery period after running, 
the Hurst exponents for healthy
and SAP subjects are moving towards each other and to the values found in the 
Pretrigger state (Pt). 

These preliminary results of the R/S analysis show a clear 
separation in the value of the Hurst exponent under controlled physical 
activity.
Further studies in larger populations are needed to confirm this result.
If the found trend would continue on larger statistics, the Hurst R/S method 
could be useful in separating SAP subjects from healthy ones, 
especially in borderline cases where clinical diagnosis
cannot be set from ECG measurement only.  

In conclusion, we have shown that fluctuations in heartbeat time series in a
controlled ergometric regimes exibit fractal properties when analysed by a
rescaled range method (Fig.4).
The rescaled range (R/S) for ergometric measurements is very well described
by the Hurst empirical law $R/S\sim n^{H}$, for $2\leq n\leq 400$, where
the Hurst exponent $H>0.5$ increases for healthy subjects, in contrast to SAP
subjects where it is found to be decreasing, during progressive physical 
activity.


\begin{thebibliography}{10}
\bibitem{ref1} C. K. Peng, S. Havlin, H. E. Stanley, and A. L. Goldberger,
Chaos {\bf 5}, 82 (1995).  
\bibitem{ref2} P. Ch. Ivanov, L. A. N. Amaral, A. L. Goldberger, 
S. Havlin, M. G. Rosenblum, Z. R. Struzik, and H. E. Stanley,
Nature {\bf 399}, 461 (1999).
\bibitem{ref3} C. L. Bolis and J. Licinio, eds.: The autonomic Nervous
System, World Health Organization, Geneva, 1999.
\bibitem{ref4} Y. Yamamoto, Y. Hoshikawa, and M. Miyashita, J. App. Phys.
{\bf 81}, 1223 (1996).
\bibitem{ref5} P. B. De Petillo, d'Armand Speers, and U. E. Ruttimann, Comp.
in Biology and Medicine {\bf 20}, 393 (1999).
\bibitem{ref6} H. E. Hurst, Trans. Am. Soc. Civ. Eng. {\bf 116}, 770 (1951).
\bibitem{ref7} J. Feder: Fractals, Plenum Press, NY, 1988,
pp. 149-183. 
\bibitem{ref8} P. Barat, N. K. Das, D. Ghose, and B. Sinha,  
Physica A {\bf 262}, 9 (1999); \\
B. Hoop, H. Kazemi, and L. Liebovitch, 
Chaos {\bf 3}, 1, 27 (1993).
\end{thebibliography}
\end{document}